# Slip-hydride interactions in Zircaloy-4: Multiscale mechanical testing and characterisation

Siyang Wang*, Finn Giuliani, T. Ben Britton

Department of Materials, Royal School of Mines, Imperial College London, London, SW7 2AZ, UK

*Corresponding author: siyang.wang15@imperial.ac.uk

## Abstract

The interactions between δ-hydrides and plastic slip in a commercial zirconium alloy, Zircaloy-4, under stress were studied using *in situ* secondary electron microscope (SEM) micropillar compression tests of single crystal samples and *ex situ* digital image correlation (DIC) macroscale tensile tests of polycrystalline samples. The hydrides decorate near basal planes in orientation, and for micropillars orientated for $<a>$ basal slip localised shear at the hydride-matrix interface is favoured over slip in α-Zr matrix due to a lower shear stress required. In contrast, for pillars oriented for $<a>$ prismatic slip the shear stress needed to trigger plastic slip within the hydride is slightly higher than the critical resolved shear stress (CRSS) for the $<a>$ prismatic slip system. In this case, slip in the hydride is likely achieved through $<110>$-type shear which is parallel to the activated $<a>$-type shear in the parent matrix. At a longer lengthscale, these results are used to inform polycrystalline samples analysed using high spatial resolution DIC. Here localised interface shear remains to be a significant deformation path which can both cause and be caused by matrix slip on planes closely-oriented to the phase boundaries. Matrix slip on planes nearly perpendicular to the adjacent hydride-matrix interfaces can either result in plastic slip within the hydrides or get arrested at the interfaces, generating local stress concentration. Through these mechanisms, the presence of δ-hydrides leads to enhanced strain localisation in Zircaloy-4 early in the plastic regime.

Keywords: Zirconium alloy; Zirconium hydride; Micropillar compression; Digital image correlation; Plasticity

# 1. Introduction

Zirconium alloys are used extensively as fuel cladding materials in water-based nuclear reactors for their high mechanical strength to neutron absorption cross section ratio and relatively good corrosion resistance. A major concern on the in-service performance of these materials, however, lies in their susceptibility to hydrogen introduction due to the water/steam operational environment and as tritium via ternary fission, rapid hydride formation as a result of the low hydrogen solubility in zirconium, and the potential detrimental effects the hydrides can have on the mechanical properties and structural integrity of the cladding material. However, the mechanical performance of these hydrides in the zirconium cladding is still not well understood. The purpose of this paper is to provide insight into dislocation-hydride interactions, for specific slip systems interacting with hydrides using micropillar testing. These tests lack external constraint, and therefore we combine this with macroscopic polycrystalline tests with high resolution digital image correlation (DIC) to understand how the hydrides will influence local behaviour. We hope these results can influence the generation of models that may be used to inform lifetime predictions for delayed hydride cracking (DHC).

The most commonly-observed hydride phase in hexagonal close packed (HCP) α-Zr is δ phase hydride. It has face centred cubic (FCC) crystal structure and forms a semi-coherent interface with the matrix [1]. The orientation relationship (OR) between α-Zr and δ-hydride is $\{0001\}_\alpha || \{111\}_\delta; <11\bar{2}0>_\alpha \,||\, <110>_\delta$ [2–6] and the habit planes of the large lens-shaped hydride packets are the $\{10\bar{1}7\}$ planes of the surrounding matrix [7,8]. Through macroscale mechanical testing on bulk specimens, it has been found that hydride precipitates can significantly degrade the ductility and toughness of bulk Zr alloys [9–14]. One potential failure mechanism is the embrittling effect of the hydrides and the propensity for hydrogen in solution to diffuse towards the crack tip at elevated temperatures together which can lead to the failure of Zr alloy fuel claddings through DHC [15–18].

However, although the hydrides normally fail through fracture, they have been found to be not purely brittle while examined at the microscale. Plastic deformation of the δ-hydride, both as a bulk [19,20] and as planar precipitates within Zr alloy matrix [21], has been routinely observed prior to fracture events during mechanical tests. This indicates that plasticity within the hydride, the matrix, and at the hydride-matrix interface may play an important role at the early stage of damage accumulation and may control how the material fails at the macroscopic scale. Furthermore, as the δ-hydride precipitates sit on $\{10\bar{1}7\}$ planes of the matrix, their relative orientation with respect to the applied load can affect the failure mode strongly in textured Zr alloys. From a micromechanical point of view, this also means that the hydrides likely have different impacts on the different slip systems of the matrix due to the anisotropy of the HCP structure. Therefore, detailed investigation into the effect of hydrides on the mechanical behaviour per slip system is required to evaluate if and how the hydrides mechanically affect the various deformation modes differently, and to analyse and predict the plastic response of the material in practical scenarios under complicated stress states.

In order to investigate the mechanical effect of the hydride precipitates on individual matrix slip systems, it is necessary to test single crystal specimens with known crystal orientations under well-defined loading conditions, such that ambiguities in the local stress states arisen from grain neighbour effects can be effectively removed. This can be achieved through *in situ* micromechanical testing, where it is possible to extract and test single crystal specimens (using geometries such as micropillars and microcantilevers) fabricated often by focussed ion beam (FIB) from polycrystalline materials with real-time characterisation [22]. Activation of target slip systems is achievable by fabricating and testing the small scale samples in specific grains based on the knowledge of the crystal orientation, the loading condition, and Schmid's law [23–27]. The application of this technique has contributed noticeably to the understanding of the mechanical properties of Zr alloys and Zr hydrides, in terms of critical resolved shear stress (CRSS) [26,28], fracture toughness [19], hydride-matrix interactions [21] and phase boundary defect evolution [27].

Industrial nuclear fuel cladding tubes are mainly made of polycrystalline fine grain Zr alloys in order to achieve high level of mechanical strength. In these materials the hydride packets preferentially decorate the grain boundaries and interconnect to form hydride stringers, and moreover, these stringers tend to align along specific texture components [3,6,29]. Since hoop stress is normally the most severe stress component that the claddings are subject to, the texture of the tubes is often controlled through processing routes such that hydrides form circumferentially upon service, in which case local mode 1 fracture events can be effectively suppressed [30,31]. However, strain-induced voids were found to form along circumferential hydrides which may lead to damage accumulation type ductile failure [32,33], though the mechanisms for the initiation of this process remain unclear.

In recent years, the development of sub-micron scale digital image correlation has enabled the detection and quantification of local deformation over relatively large areas on deformed material surfaces [34–42]. This technique is deemed appropriate for addressing the early stage initiator of damage accumulation type failure due to its high sensitivity to small deformation and high spatial resolution. Here we employ this technique to reveal the evolution of strain distribution within and around circumferential hydrides in Zircaloy-4 upon plastic deformation, which may help better understand the local deformation behaviour at and near the hydride precipitates.

In this paper, *in situ* scanning electron microscope (SEM) micropillar compression tests and *ex situ* DIC macroscale tensile tests are utilised to study the interactions between plastic slip in α-Zr and δ-hydrides in single and polycrystalline Zircaloy-4 specimens, respectively. The paper is organised as follows. Section 2 gives a description of the sample preparation methods and details about the mechanical tests. The experimental results are shown in section 3, which are then discussed in section 4 in terms of the anisotropic effects of the hydrides on the matrix slip behaviour along with the implications of these effects in the initiation of hydride-induced failure of Zr alloys.

## 2. Experiments

### 2.1. Micropillar compression tests

A cuboid (approx. 10 mm × 10 mm × 2 mm) fine grain Zircaloy-4 (Zr-1.5%Sn-0.2%Fe-0.1%Cr in weight% [43], with an average grain size ≈ 11 μm) sample was cut from a rolled and recrystallised plate and heat treated for 336 h at 800 °C in Ar atmosphere in order to produce large 'blocky alpha' grains with an average size >200 μm [44]. Electrochemical hydrogen charging of the sample was carried out in 1.5 wt% sulfuric acid at 65 °C for 24 h using a current density of 2 kA/m$^2$ [45,46], followed by an annealing at 400 °C for 5 h to homogenise the hydrogen distribution. The sample was then cooled down to room temperature with a cooling rate of 1 °C/min to promote the formation of δ phase hydrides. The microstructure of the sample after these treatments is shown in Figure 1, where intragranular, intergranular and twin boundary hydrides can be observed.

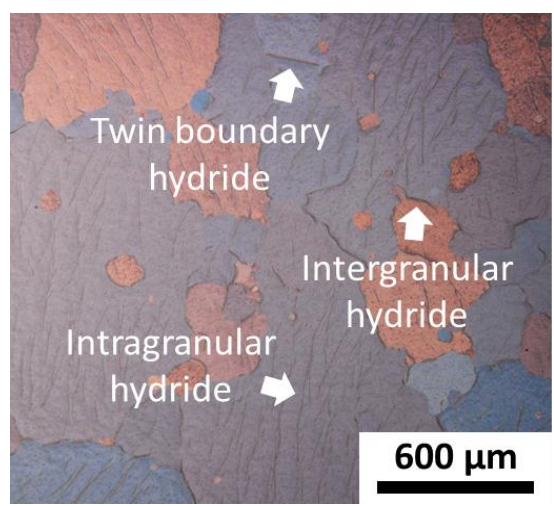

Figure 1 Polarised light optical micrograph of hydrided large grain Zircaloy-4.

The sample was mechanically polished with colloidal silica and then electropolished with 10 vol% perchloric acid in methanol at -40 °C for 90 s under an applied voltage of 30 V.

Electron backscatter diffraction (EBSD) was used to choose crystal orientations required to fabricate and test micropillars in specific grains such that target slip systems can be activated while the others are switched off.

EBSD characterisation was carried out on a FEI Quanta 650 SEM equipped with a Bruker eFlashHR (v2) EBSD camera using 20 kV electron beam voltage. Within this paper, all the unit cell representations of crystal orientations are plotted with respect to the viewing angle of the images they are attached to.

Micropillar fabrication was achieved using Ga focussed ion beam (FIB) on a FEI Helios Nanolab 600 Dualbeam microscope, using an acceleration voltage of 30 kV and beam currents from 21 nA to 1 nA. High beam currents were used to mill 50-μm-diameter circular trenches around the micropillars for *in situ* visualisation of the pillars during compression tests, while low beam currents enabled the final tailoring of the pillar contour. Square micropillars with 5 μm mid-height width, 15 μm height, and 2° taper angle were fabricated within two grains – one has a high Schmid factor for $<a>$ basal slip and another has a high Schmid factor for $<a>$ prismatic slip, assuming a uniaxial stress applied along the direction perpendicular to the sample surface. These two grains will hence be termed grain 1 (where $<a>$ basal slip is favoured) and grain 2 (where $<a>$ prismatic slip is favoured) respectively herein. For the micropillars in these two grains, the calculated highest Schmid factors for the available slip systems of Zircaloy-4 in the micropillar compression geometry are listed in Table 1. Due to their significantly higher CRSS values [28], the pyramidal slip systems are unlikely to be active even though the Schmid factors for some of them are relatively high.

Table 1 The highest Schmid factors for all potential slip systems of the Zr micropillars under a uniaxial stress applied perpendicular to the sample surface.

|  | $<a>$ basal | $<a>$ prismatic | $<a>$ pyramidal | First order $<c+a>$ pyramidal |
|---|---|---|---|---|
| Grain 1 | 0.47 | 0.29 | 0.43 | 0.35 |
| Grain 2 | 0.12 | 0.49 | 0.47 | 0.48 |

Some of the micropillars in these two grains were fabricated site-specifically such that they contain intragranular hydride packets. Since the $\{10\bar{1}7\}$ habit planes of the hydrides are close to the $\{0001\}$ basal planes of the matrix, in grain 1 where the c-axis (basal plane normal) is ~45° inclined, the hydride packets make a similar angle to the pillar top surface as shown on the SEM image in Figure 2(a), where a lens-shaped hydride packet which appears as a darker contrast region than the matrix (the hydride packet penetrates the pillar from its top left corner to its right edge, and the hydride-matrix interfaces at the right edge of the pillar are highlighted with black lines) can be observed on the pillar front surface. Pillars in grain 2, where the c-axis is nearly parallel to the pillar top surface, contain hydride packets that sit nearly vertically, as shown on the SEM image in Figure 2(b), where the hydride-matrix interfaces at the top edge of the pillar front surface are highlighted with black lines.

*In situ* micropillar compression tests were carried out using a displacement-controlled Alemnis nanoindenter on the FEI Quanta SEM. Micropillars were compressed with a 10 µm-diameter circular flat punch indenter in order to achieve a (near) uniaxial stress state. All the pillars were deformed to a displacement of 1.5 µm, corresponding to a nominal strain of 10%. The loading and unloading speeds used were 5 nm/s and 50 nm/s, respectively (i.e. a nominal loading strain rate of ~3x10$^{-4}$ s$^{-1}$).

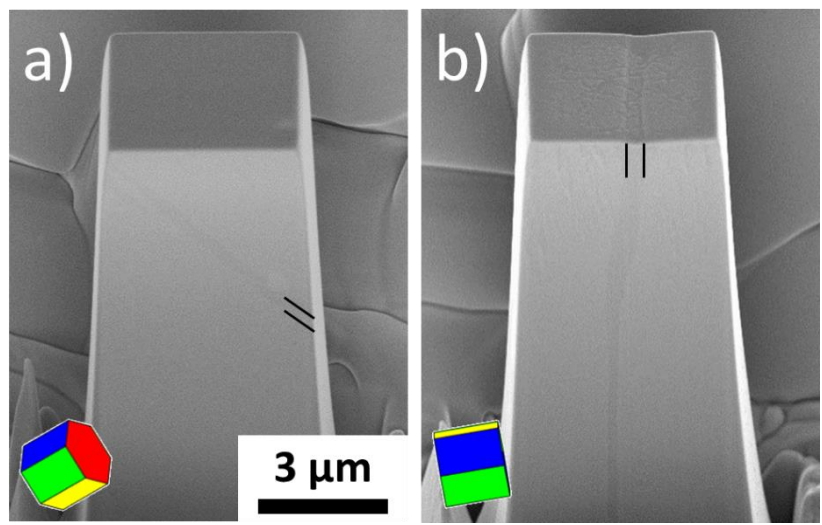

Figure 2 SEM images of hydride-containing micropillars in (a) grain 1 and (b) grain 2, with inserts of unit cell representations of α-Zr matrix crystal orientations, oriented with respect to the viewing angle. The black lines highlight the positions of the hydride packets. The front face of these pillars is inclined at 52° to the viewing angle.

## 2.2. Macroscale tensile tests

For the macroscale tensile tests, a 'dogbone'-shaped sample was cut from the as-received rolled plate in such a manner that the loading direction of the sample is aligned with the plate rolling direction (RD), as shown in Figure 3. The material has a 'split basal' texture (Ref. [44]) with basal poles oriented ~30° away from the plate normal direction (ND) towards the plate transverse direction (TD). The sample was then charged with hydrogen using the method introduced in section 2.1 to ~240 ppm-wt [H], determined through extracting the terminal solid solubility on dissolution (TSSd) using differential scanning calorimetry (DSC, Mettler Toledo DSC1) and calculations according to Ref. [47]. The geometry and microstructure of the sample are shown in Figure 3. In the fine grain material, the hydrides form interconnected hydride stringers (as per Ref. [6]). The way the sample was machined promoted the formation of hydride stringers along the loading direction in the present configuration, and this was chosen to be similar to the practical case of fuel claddings where circumferential hydrides are subject to hoop stress.

After mechanical polishing, the tensile test sample was polished with broad Ar ion beam in a Gatan PECS II System for EBSD characterisation. Details about the EBSD analysis can be found in section 2.1. Two-dimensional strain mapping within a region of interest (ROI) on the sample surface that contains a hydride stringer was achieved using DIC. A speckle pattern with an average speckle size of ~40 nm was applied to the sample surface *via* vapour-assisted gold remodelling adapted from those introduced in Ref. [34–36]. In particular, we opt to heat up the styrene to about 60 °C in order to significantly increase the vapour pressure and therefore speed up the remodelling process.

A displacement-controlled Gatan MTest 2000E mechanical testing stage was used for the tensile tests. The tests were performed under displacement-controlled mode (crosshead speed = 0.033 mm/min) in a series of increasing displacement increments and *ex situ* DIC strain mapping was carried out after each step of deformation. Images of the speckle pattern within the DIC-ROI were taken on a Zeiss Sigma 300 SEM. A matrix of 6×7 images were taken with 25% overlap, and each of the images has a resolution of 3072×2304 pixels where the pixel size is ~6 nm. Stitching of the SEM images were carried out using the stitching plugin [48] in the Fiji [49] software package based on ImageJ [50]. Subpixel stitching accuracy was achieved through linear interpolation and the intensities in the overlapped regions were blended linearly. The size of the total composite frame (the entire DIC-ROI) is ~80×75 μm$^2$.

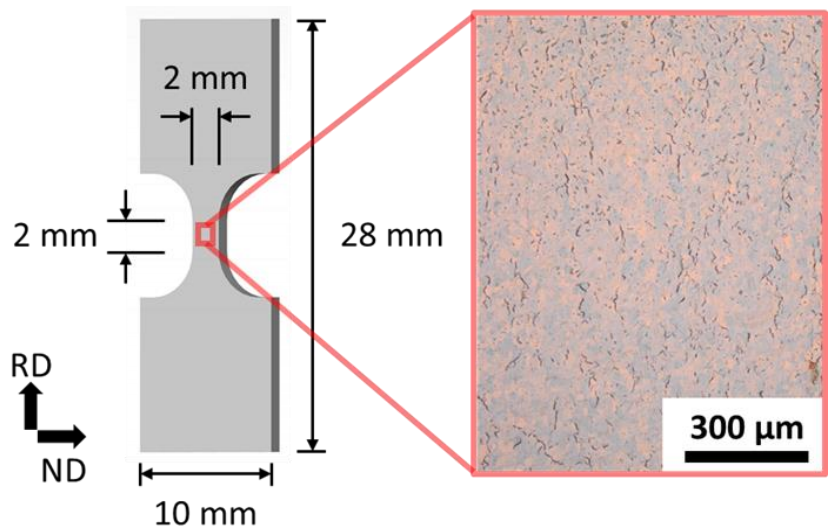

Figure 3 Geometry and microstructure of the hydrogen charged fine grain Zircaloy-4 specimen for macroscale tensile tests. RD and ND denote rolling direction and normal direction (of the rolled plate), respectively.

DaVis 8.4 software was utilised for the image correlation and strain estimation, with a subset size of 21x21 pixels and a step size of 5 pixels. This gives a strain spatial resolution of ~30 nm. Within the present paper, all the DIC strain maps are projected into the undeformed configuration, in order to carry out slip trace analysis using the EBSD results obtained pre-testing.

## 3. Results

### 3.1. Micropillar compression tests

#### 3.1.1. $<a>$ basal slip and near parallel hydrides

Compression tests were first carried out on two as-received (no hydride present) micropillars, and it was found that they exhibited similar slip behaviour. An SEM image of one of these as-received pillars after compression testing is shown in Figure 4(a), with an insert of the unit cell representation of the crystal orientation for all the pillars in grain 1. A slip band can be observed on the front surface of the pillar, between its top left corner and right edge. The dashed line overlaid on the SEM image shows the theoretical orientation of the $<a>$ basal slip band derived from the crystal orientation of the pillar measured by EBSD. The parallel between the observed slip band on the pillar and the dashed line suggests that $<a>$ basal slip is the dominating deformation mode in this pillar, in agreement with the Schmid factor analysis shown in section 2.1.

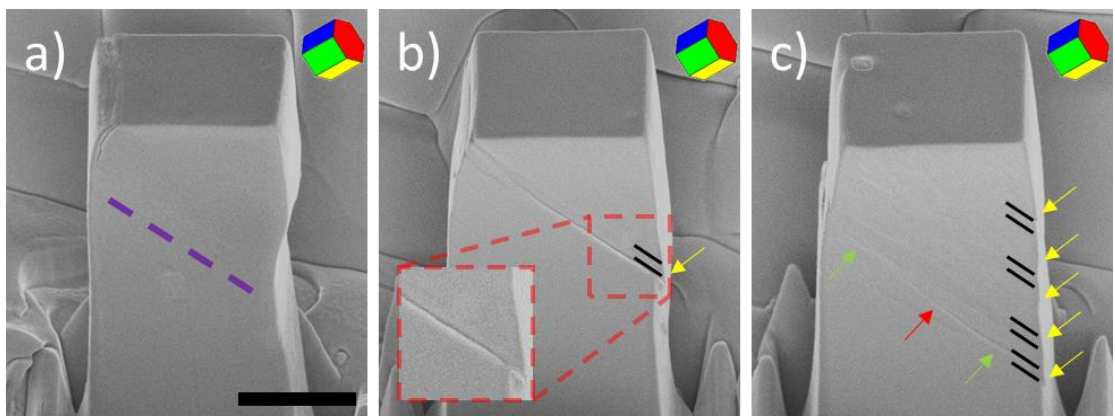

Figure 4 Post-deformation SEM images of (a) as-received and (b,c) hydride-containing micropillars in grain 1, with inserts of unit cell representations of matrix crystal orientations and black lines highlighting the hydride packets. The pillar in (b) contains a single hydride packet while the one in (c) contains multiple hydride packets. The scale bar is 3 µm long.

The critical resolved shear stress (CRSS) for $<a>$ basal slip in Zircaloy-4 can be extracted from the engineering stress-strain responses of the as-received micropillars (Figure 5), through multiplying the yield stress (obtained using the 0.2% offset method) by the calculated Schmid factor for $<a>$ basal slip. From the two pillars tested, measured yield stresses of 314 MPa and 325 MPa, with a Schimd factor of 0.47 give a calculated CRSS for $<a>$ basal slip of 148 MPa and 153 MPa.

Compression tests were then carried out on micropillars fabricated also within grain 1 but containing hydride packets. Figure 4(b) shows the post-deformation SEM image of the hydride-containing micropillar shown in Figure 2(a). Instead of following the trace of $<a>$ basal plane as the slip bands on the as-received pillars do, the shear band formed on this hydride-containing pillar is found close to the lower interface between the hydride packet and the matrix (the hydride-matrix interfaces are highlighted with black lines on the image). Moreover, this shear band formed at the hydride-matrix interface is more localised than that formed on the as-received pillar, and this localised shear band has led to the formation of a distinct step at the right edge of the pillar, as marked with a yellow arrow in Figure 4(b).

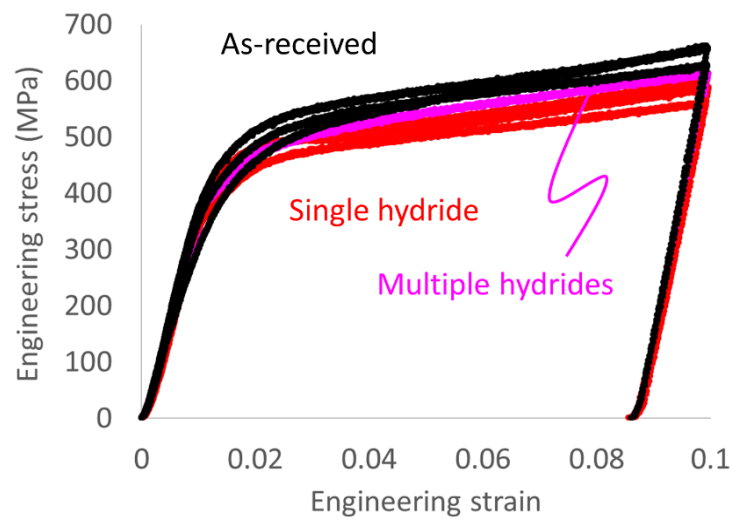

Figure 5 Engineering stress-strain curves for the as-received and hydride-containing micropillars in grain 1, oriented to activate $<\boldsymbol{a}>$ basal slip in the α-Zr matrix.

The post-deformation SEM image of another type of hydride-containing pillars in grain 1 is shown in Figure 4(c). Unlike the micropillar shown in Figure 4(b) that contains a single hydride packet, multiple hydride packets lying parallel to each other can be seen on the front surface of this pillar, and the interfaces between the hydride and the matrix are highlighted with black lines at the right edge of the pillar on the image. Similar to the shear band observed in the pillar that contains a single hydride packet, some of the shear bands formed on the front surface of this pillar are also observed near the hydride-matrix interfaces. These shear bands have also resulted in the formation of small steps at those points where the hydride-matrix interfaces meet the right edge of the pillar, as highlighted with the yellow arrows in Figure 4(c). Apart from these, another distinct shear band, as highlighted with the red arrow on the image, is present away from any of the hydride-matrix interfaces but seems to have interconnected with the two shear bands at the hydride-matrix interfaces that have been marked with green arrows. This interconnecting shear band aligns closely to the trace of the {0001} basal plane on the pillar front surface (see the dashed line in Figure 4(a)).

Figure 5 summarises the engineering stress-strain responses of all the micropillars within grain 1 (oriented to activate basal slip) that have been tested. The flow stresses (the stress level at a particular strain upon plastic deformation) for the hydride-containing pillars are in general ~10% lower than those for the as-received pillars, while the strain hardening rates (determined from the engineering stress-strain responses using the method introduced in Ref. [26]) for all these pillars are comparable values. Amongst all the hydride-containing pillars, the flow stresses for the one with multiple hydrides are the highest (see the pink curve in Figure 5).

### 3.1.2. $<a>$ prism slip selectively blocked by hydrides

Micropillars were tested in a second grain (grain 2), selected to activate $<a>$ prism slip and have hydrides that run vertical in the micropillars. Post-deformation SEM micrographs indicate that all five as-received pillars in this grain show similar deformation behaviour. An example of the appearance of the as-received pillars after deformation is shown in Figure 6(a). Discrete slip bands can be observed on the front surface of the pillar which align with the trace of one of the prismatic crystallographic planes (the yellow-coloured plane on the unit cell representation, trace plotted with the dashed line overlaid on the image). This indicates that $<a>$ prismatic slip is the dominating deformation mode for the as-received pillars in grain 2, in line with the Schmid factor analysis. The distribution of slip bands on the as-received pillars in grain 2, however, is evidently different from that for the as-received pillars in grain 1 where the $<a>$ basal slip is activated (see Figure 4(a)). For the pillars in grain 2, narrow slip bands can be seen throughout the pillar body from top to bottom. However, for the pillars in grain 1 there is only one relatively broad slip band per pillar observed close to the top of the pillars (this is similar to observations in titanium [25]). Figure 6(b) shows an SEM image of the same pillar as that shown in Figure 6(a) but viewed from its 'left' (the left surface of the pillar in Figure 6(a)). This image shows the individual slip steps formed on the side surface of the pillar which, as expected, also align well with the trace of the yellow-coloured prismatic plane on the unit cell representation (the dashed line on the image).

The CRSS for $<a>$ prismatic slip can hence be extracted from the engineering stress-strain responses of the pillars tested, using the same method as introduced previously. For the five tested pillars, the measured CRSS values are 105 MPa, 112 MPa, 101 MPa, 123 MPa and 112 MPa, respectively.

As introduced previously, the hydride packets lie close to the matrix {0001} basal plane. Therefore, viewing the pillars in grain 2 from a side surface which is nearly parallel to the c-axis of the HCP structure allows for the observation of the cross-section of the hydride packets and therefore their impacts on the deformation behaviour. Figure 6(c,d,e) show the post-deformation SEM images of three different hydride-containing pillars viewed from the same direction (as that in Figure 6(b) for the as-received pillar), with black lines highlighting the hydride-matrix interfaces near the top edges of the pillars.

In Figure 6(c), some distinct slip bands can be observed close to the bottom of the pillar. These slip bands, unlike those observed on the as-received pillars which have extended across the entire pillar body, have apparently been affected by the vertical hydride packet that sits in the middle. The three slip bands in the matrix that have been marked with yellow arrows in the figure seem to have been arrested by the hydride packet as no obvious deformation feature is observable on the other side of the hydride. For the slip band marked with the red arrow, on the other hand, a faint slip band can be seen on the other side of the hydride (marked with the green arrow) and these two bands align with each other and are interconnected at the hydride packet. This implies that the original slip band which likely initiated within the matrix (but not necessarily at the edge of the pillar) was able to trigger slip in the hydride and then in the matrix on the other side during the deformation process, in agreement with observations in macroscale tests [51]. Besides, it is unlikely that slip started within the hydride and propagated out, based on the observation of substantially higher number of slip bands in the matrix than in the hydride, as well as the blocked slip bands in the matrix (rather than in the hydride) at the hydride-matrix interfaces.

The pillar in Figure 6(d) contains a hydride packet of similar thickness to the one shown in Figure 6(c). One slip band on this pillar (as marked with the red arrow) can be seen to have extended across the entire pillar, shearing the hydride packet too. Besides, as shown in the zoomed in region within Figure 6(f), a few other slip bands on this pillar (marked with green arrows) have also propagated through the hydride, but, these slip bands seem to have deviated slightly away from their original direction while going through the hydride, as local kinks (marked with purple arrows) can be observed within the hydride in those regions where the slip bands intersect the hydride packet.

The pillar in Figure 6(e) contains a hydride packet which is ~1/3 the thickness of those hydride packets in the two pillars shown above (Figure 6(c,d)). Although this hydride packet is remarkably thinner than the previous ones, it also poses an impact to the slip behaviour of the pillar. For instance, the slip band marked with the yellow arrow has not propagated through the hydride packet, while those marked with the two red arrows seem to have managed to trigger slip in the hydride, although the faint slip bands observed on the other side of the hydride suggest that there may have been some interruption from the hydride as the original slip bands propagated through.

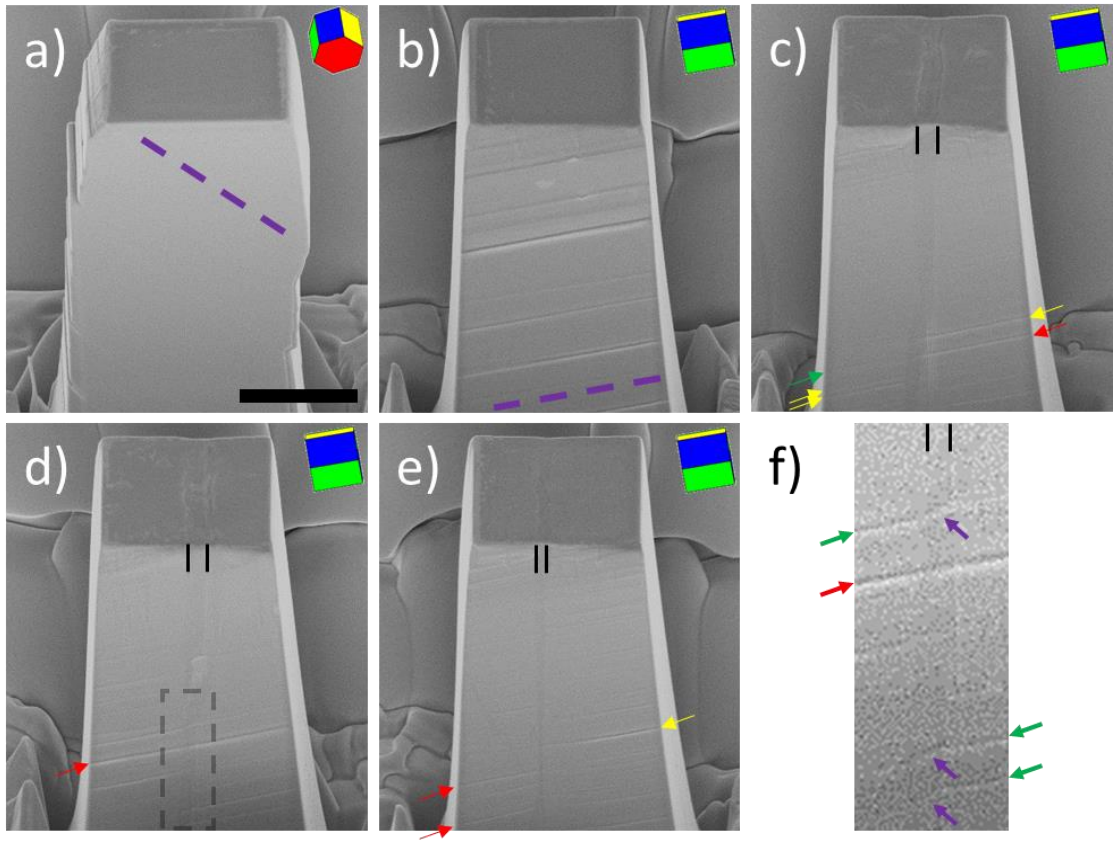

Figure 6 Post-deformation SEM images of (a,b) as-received micropillar ((b) is the front view of the pillar left surface in (a)) and (c,d,e) hydride-containing micropillars in grain 2, oriented to activate $<a>$ prism slip in the α-Zr matrix. With the figures include inserts of unit cell representations of matrix crystal orientations and black lines highlighting the hydride packets. (f) magnified image of the region within the dashed box in (d). The scale bar is 3 µm long.

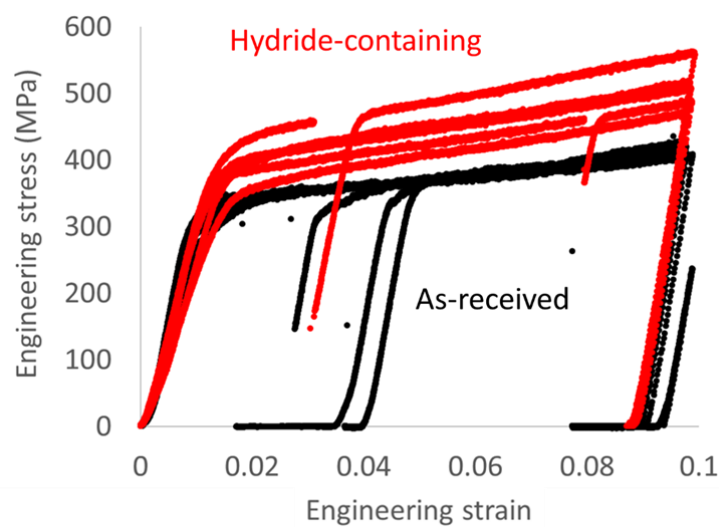

Figure 7 Engineering stress-strain curves for the as-received and hydride-containing micropillars in grain 2, oriented to activate $<a>$ prism slip in the α-Zr matrix.

The engineering stress-strain responses of all the pillars tested in grain 2 are summarised in Figure 7, with the ones for the as-received pillars plotted in black and those for the hydride-containing pillars plotted in red. The two sets of data exhibit three major differences:

i. The yield stresses are generally higher for the hydride-containing pillars than the as-received pillars;

ii. The strain hardening rates (determined using the method in Ref. [26]) are generally higher for the hydride-containing pillars than the as-received pillars;

iii. The hydride-containing pillars exhibit less stress drops than the as-received pillars upon plastic deformation. Some of the stress drops on the stress-strain curves for the as-received pillars are so significant that the stress dropped to zero during the testing processes. The origin of these stress drops observed will be discussed in Section 4.2.

### 3.2. Macroscale tensile tests

The engineering stress-crosshead displacement curve for the hydride-containing fine grain tensile specimen is plotted in red in Figure 8. For comparison, the mechanical response of a same-sized as-received fine grain Zircaloy-4 sample tested under the same conditions is plotted in black. The yield stress and work hardening rate for the hydride-containing sample are lower than those for the as-received sample. DIC strain mapping of the ROI on the hydride-containing sample surface was carried out within the plastic deformation regime at those three points marked with blue crosses on the curve, and the corresponding maximum shear strain maps of the DIC-ROI are given in Figure 9. The grain and phase boundaries in this region are overlaid on the strain maps, with grain boundaries plotted in white and phase boundaries in yellow. Note that the hydride structure within the DIC-ROI chosen is indicative of one of the hydride stringers observed in the optical microscope image in Figure 3, and it is revealed by high magnification SEM imaging here that the hydride stringers are consist of smaller individual hydride packets which agrees with our previous EBSD study [6]. It is also worth noting that although some of these hydrides are not interconnected on the surface examined, they may be linked to each other in the 3D network based on the formation mechanisms proposed in Ref. [6].

Generally, the distribution of the maximum shear strain in the α-Zr matrix in those regions away from the hydrides is similar to that reported in the literature for the same material, where relatively diffuse and homogeneous slip as well as grain boundary shear accommodate the imposed strain [39]. For those regions within and around the hydrides, there are in general three types of distinct features that can be seen:

i. Significantly lower level of maximum shear strain within the hydrides than that within the matrix;

ii. Localised shear along some of the hydride-matrix interfaces, as marked with green arrows on the maximum shear strain maps.

Arrested slip bands at most of the hydride-matrix interfaces, as marked with red arrows on the maximum shear strain maps.

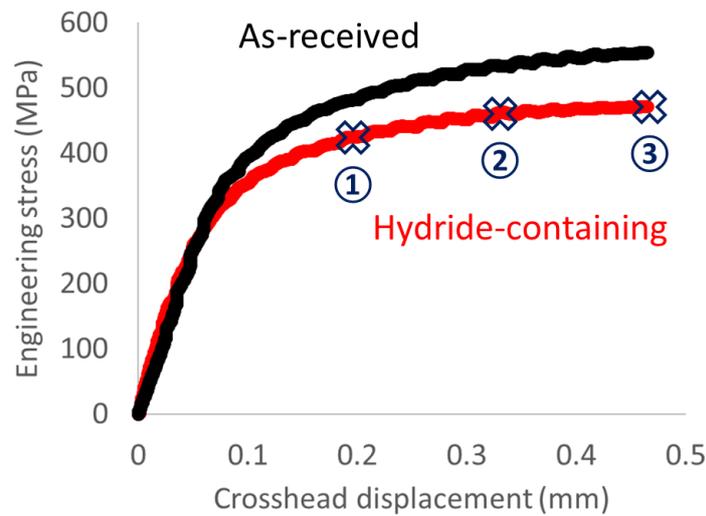

Figure 8  Engineering stress-crosshead displacement curves for the as-received and hydride-containing macroscale tensile specimens.

It is worth noting that in Figure 9 only deformation features at those hydride-matrix interfaces that do not coincide with the matrix grain boundaries are highlighted, and therefore the presence of these features is due likely to the hydrides in the microstructure, but not to the effect of grain boundaries on the slip behaviour. It is evident that the hydride-matrix interfaces along which localised shear can be observed (those marked with green arrows on the maps) are often inclined from the loading direction for roughly 45°, while those where the slip bands in the matrix were arrested (marked with red arrows on the maps) do not have such character. In other words, the slip bands in the matrix get arrested easily once they reach the hydride-matrix interfaces, regardless of the geometry of the local microstructure.

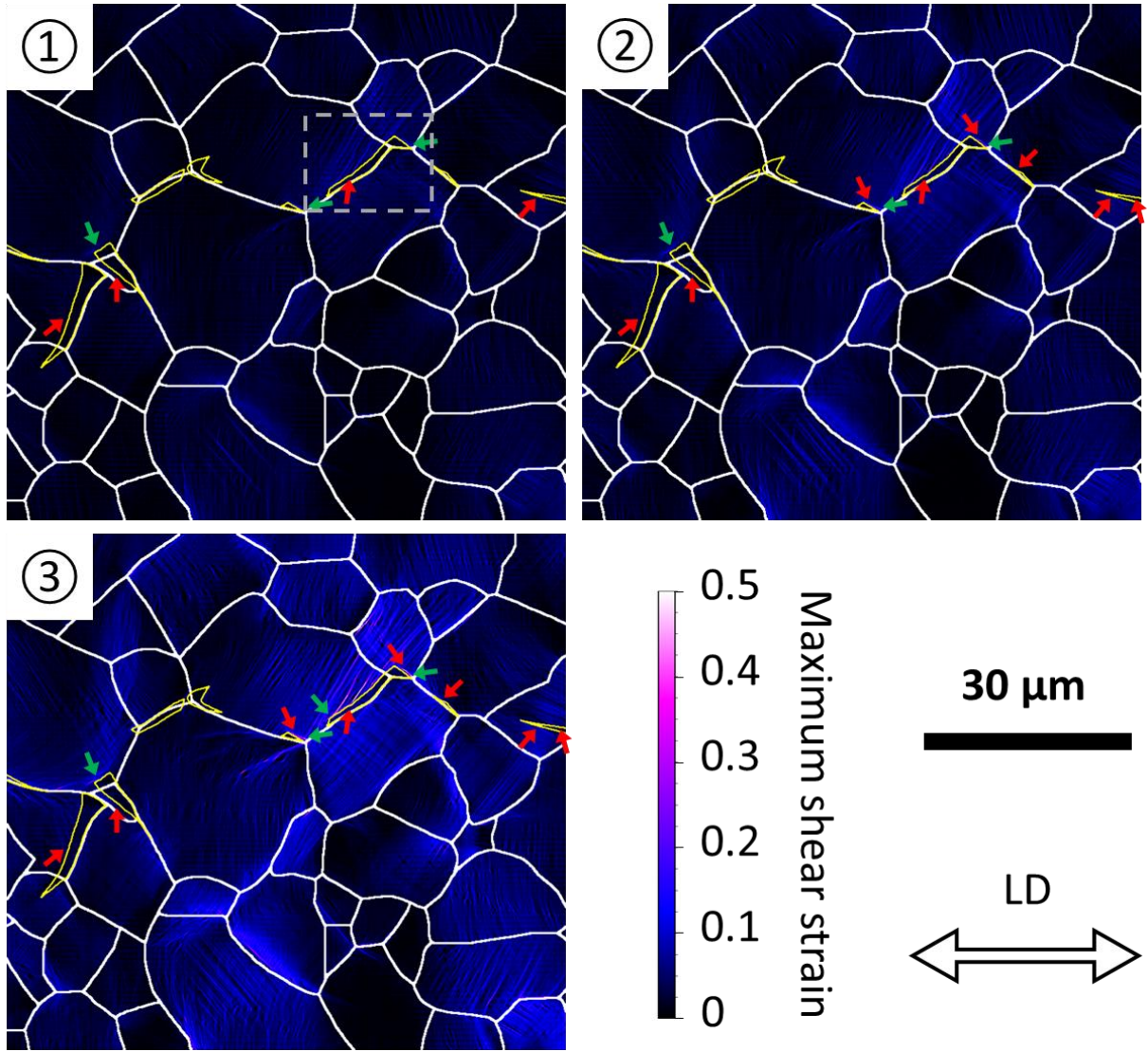

Figure 9 Maximum shear strain maps of the DIC-ROI during plastic deformation of the tensile specimen. Maps ①, ② and ③ were obtained after unloading from the three points marked with the crosses on the stress-displacement curve in Figure 8 correspondingly. The grain and phase boundaries in this region are also shown, with grain boundaries plotted in white and phase boundaries in yellow. LD denotes loading direction.

## 4. Discussion

### 4.1. $<a>$ basal and $<a>$ prismatic slip in Zircaloy-4 micropillars

In the present work, the measured CRSS values for $<a>$ basal and $<a>$ prismatic slip in α-Zr through micropillar compression tests are 150 MPa and 110 MPa, respectively. For these two slip systems in α-Zr, CRSS values reported in the literature are 204 ± 66 MPa for $<a>$ basal slip and 153 ± 30 MPa for $<a>$ prismatic slip and these results are obtained through microcantilever bending tests and fitting with a crystal plasticity model [28]. The two sets of results are in relatively good agreement though broadly the CRSS values measured here using pillar compression geometry are moderately lower than those measured from cantilever tests. This might arise from the difference in the materials used (the cantilevers tested in Ref. [28] were made of commercially pure Zr rather than Zircaloy-4, likely with a different oxygen content) and/or the size/geometry effect in small scale mechanical tests. The ratios of CRSS values for $<a>$ basal to $<a>$ prismatic slip are both ~1.3 for the two sets of results. It is also worth noting that the CRSS for $<a>$ basal slip measured here agrees well with our previous measurement using the same method (micropillar compression) but with pillars having a slightly different crystal orientation [26].

Plastic deformation is accommodated by $<a>$ basal and $<a>$ prismatic slip for pillars in grain 1 and 2, respectively. However, unlike the diffuse character of $<a>$ basal slip (exhibited as the broad shear band that has formed on the side surface of the pillars in grain 1, Figure 4(a)), $<a>$ prismatic slip appears as discrete and localised slip bands as can be observed on the side surface of the pillars in grain 2, Figure 6(a,b). The highly planar character of the slip bands on the pillars in grain 2 implies that the strain imposed during the tests is accommodated by $<a>$ prismatic slip with negligible cross-slip of dislocations onto other crystal planes. This is also supported by the Schmid's law where for these pillars the Schmid factor for the $<a>$ basal slip system that shares an $<a>$ direction with the activated $<a>$ prismatic slip system is only 0.09. For the pillars in grain 1, on the other hand, the broad/diffuse slip character is deemed an indication of cross-slip, presumably onto the prismatic planes since pyramidal slip was found to be rather difficult in Zr [52] (the CRSS is 532 MPa [28]). For the pillars in grain 1, the Schmid factor for the $<a>$ prismatic slip system that shares an $<a>$ direction with the activated $<a>$ basal slip system is 0.18. For this grain and loading state, the ratio of the Schmid factors for $<a>$ basal to $<a>$ prismatic slip is 2.6 while the ratio of CRSS being 1.3, indicating that this $<a>$ prismatic slip system, although not being the most favourably oriented, is under noticeable amount of stress upon deformation and this may have made it easier for cross-slip from basal to prismatic plane to occur. For uniaxial tests, due to the geometry of the HCP crystal structure it is possible to activate $<a>$ prismatic slip while switching off $<a>$ basal slip, however when $<a>$ basal slip is well aligned, the cross-slipping $<a>$ prismatic slip system necessarily has a Schmid factor that is not near-zero. Therefore, apart from the intrinsic properties of these two slip systems, this geometrical effect may have also partly contributed to the routinely observed experimental phenomenon that basal slip in Zr is often accompanied by prismatic slip [52].

## 4.2. Effect of hydrides on the plastic deformation of Zircaloy-4

Unlike the as-received pillars which deform by $<a>$ basal slip, the hydride-containing pillars in grain 1 were found to shear locally at the hydride-matrix interfaces under stress and this type of shear can occasionally be accompanied by local $<a>$ basal slip (Figure 4(b,c)). It is evident on the stress-strain curves (Figure 5) that this interface shear (which is potentially coordinated slip) requires a lower applied stress to activate than that for $<a>$ basal slip in the as-received pillars, which is likely the reason for the observed change in deformation path upon the introduction of hydride packets. This can be supported by the observation that, for the pillar that contains multiple hydride packets where $<a>$ basal slip is operative alongside the interface shear, the flow stresses are higher than those for the pillars with only one hydride packet where the deformation was accommodated by interface shearing alone.

The slip band structures formed on the hydride-containing pillars in grain 2 indicate that while the pillars were deformed into the plastic regime, plastic slip ($<a>$ prismatic) took place in the matrix and, when the slip bands reached the hydride-matrix interfaces, the hydride packets can retard the propagation of the slip bands. In general, this effect of the hydrides on matrix slip does not seem to vary strongly with the thickness of the hydride packets (Figure 6(c,d,e)).

Propagation of slip bands through the hydrides is also observed in a few cases, which is likely achieved through crystallographic slip of the hydrides since well-defined straight slip bands were routinely observed within the hydride packets. As these slip bands in the hydrides are not parallel (and nearly perpendicular) to the basal plane of the surrounding matrix, they are not likely due to slip within the Zr ligaments in between the microscale hydride platelets (as observed by Weekes *et al.* [21]), but necessarily deformation of the hydride itself. The difference between the flow stresses for the as-received and hydride-containing pillars in grain 2 is evident but not significant (Figure 7, the difference is ~40-90 MPa upon yielding and ~50-130 MPa at 10% strain), indicating that the CRSS for plastic slip in the hydride is not significantly higher than that for $<a>$ prismatic slip of the matrix, even though the dislocation density is generally higher in the hydride than in the matrix [6] and this high dislocation density may result in extra hardening.

Due to the OR between the hydride and the matrix, there are only two possible hydride orientations (which are twins to each other and share a common {111} crystal plane) for a single matrix crystal and these two orientations often co-exist within a single hydride packet due to the misfit strain present [6]. For the pillars in grain 2, EBSD analysis on the sample top surface revealed that the two possible hydride crystal orientations are both present. These two orientations are shown by the unit cell representations in Figure 10, in the frame of Figure 6(f) which has also been shown here.

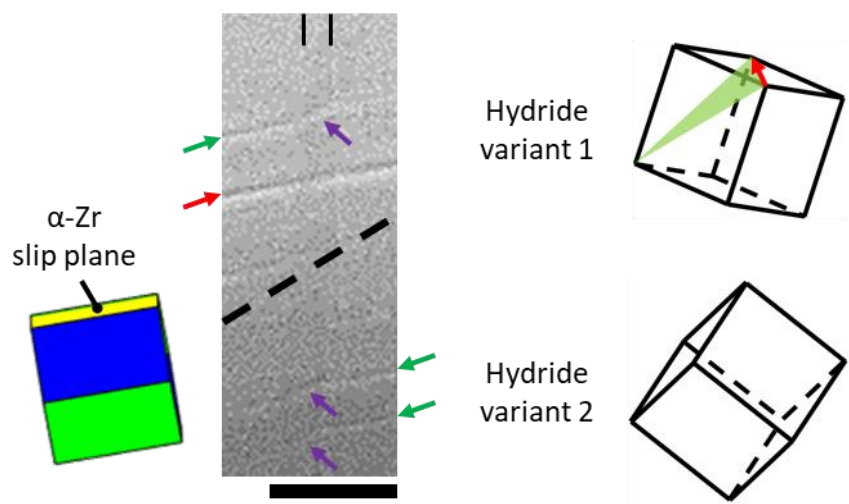

Figure 10 Slip band structure within and around a hydride packet on a micropillar (Figure 6 (d,f)), with unit cell representations of the crystal orientations of the two hydride variants and black lines highlighting the hydride packets. The dashed line overlaid on the image is the trace of the $\{111\}$ plane plotted in green in the unit cell representation of hydride variant 1 on the pillar front surface. The scale bar is 1 μm long.

The crystal structure of δ-hydride is analogous to that of calcium fluoride ($CaF_2$), with the hydrogen atoms occupying the tetragonal interstitial sites of the Zr FCC lattice. Common ionic crystals with this type of structure also include $UO_{2(+x)}$ and $ZrO_2$. The only observed slip system in bulk polycrystalline δ phase Zr hydride is $\{111\}<110>$ [20]. In the present case, the initiation of plastic deformation occurred first in the α-Zr matrix (based on the observation of the slip band structure on the pillars, in agreement with the literature that the hydride is generally stronger than the matrix [19]) and plastic slip within the hydride took place as the slip bands in the matrix reach the phase boundaries. In another word, the onset of plasticity in the hydrides is a result of the $<a>$-type shear in the matrix and dislocations reaching the interface. Since a $<110>$ direction of the hydride is parallel to the active $<a>$ ($<11\bar{2}0>$) direction of the matrix due to the hydride-matrix OR, it is likely that the slip within the hydride occurred in the $<110>$ direction that is parallel to the $<a>$ direction of the activated matrix slip system. Besides, the slip bands in the hydride that have been marked with purple arrows in Figure 10 are parallel to the dashed line overlaid on the image, which is the trace of the $\{111\}$ plane plotted in green in the unit cell representation of hydride variant 1. This suggests that slip in the hydride likely occurred on this plane. Furthermore, the $<110>$ direction of the hydride phase that is parallel to the $<a>$ slip direction of the matrix lies within this specific $\{111\}$ plane, further supporting the potential mechanism that $<a>$-type shear in the matrix has triggered $<110>$-type shear in the hydride in the same direction and on the associated OR related crystal plane. Moreover, slip events within the hydride took place on this specific $\{111\}$ plane but not on the other crystal planes which also have the active $<110>$ slip direction within them, presumably because the $\{111\}$ plane for the active system is more closely-oriented to the matrix slip plane than the others and enables shear to be propagated through the structure.

The shear band marked with the red arrow in Figure 10, however, is not parallel to the trace of any of the {001}, {011} or {111} planes for the two hydride variants. Presumably, this is due to either that this band is broader than the others so the change in its direction while propagating through the hydride is not readily observable on the SEM image, or that the formation of this specific shear band within the hydride is a result of more complicated deformation mechanisms than the activation of a single slip system. Note that non-crystallographic slip has been previously observed during deformation of $UO_{2(+x)}$ at high temperatures [53–56] and this was attributed to the cross-slip of screw dislocations from {111} planes to {001} or {011} planes. Similar mechanisms might have operated here for δ-hydride and gave rise to the non-crystallographic apparent slip plane observed macroscopically.

There are three possible reasons that may account for where some of the slip bands in the matrix are observed to not have evidently crossed the hydride-matrix interface:

i.  The (out of plane) shear strain within and on the other side of the hydride packet is not high enough for the slip bands to be detectable on the SEM images;
ii. The local shear stress at the front of the incoming slip bands is not high enough to induce shear within the hydrides;
iii. The local crystal orientation of the hydride is variant 2 in Figure 10 rather than variant 1, making slip in the hydride more difficult to occur as the potential slip planes for variant 2 which contain the $<110>$ direction parallel to the matrix slip direction are not as spatially close to the matrix slip plane as the {111} slip plane for variant 1.

For the pillars in grain 2, the increased yield stresses for the hydride-containing pillars as compared to the as-received pillars indicate that the vertically-located hydride packets have promoted the strength of the pillars in general, and a higher level of externally applied stress is required to initiate plastic deformation. The inhibition of plastic deformation in the matrix, due to the hydride packets, has led to an increase in the strain hardening rate for the hydride-containing pillars as the hydrides evidently act as barriers for the propagation of slip in the matrix. Interestingly, the strain hardening rates determined from the engineering stress-strain responses for all the hydride-containing pillars are of a similar value while the hydride thickness varies from pillar to pillar. This along with the negligible variation of the retarding effect of the hydrides on matrix slip with hydride thickness (Figure 6 (c,d,e)) suggest that the change in hardening rate and the influence on the plastic flow due to the presence of the hydride packets originate from the differences in the intrinsic properties of the hydride and the matrix, and are insensitive to the thickness of the hydrides. This would be consistent with a dislocation pile-up at the hydride-matrix interface, which is reduced only when slip propagates through the interface back to a lower level and then a new dislocation joins the pile-up to increase the flow stress again. These findings also imply that plastic slip in the hydrides progresses relatively smoothly once it initiates at the phase boundary.

The stress drops on the engineering stress-strain responses of the as-received pillars are presumably due to abrupt separation between the indenter tip and the pillar top surface during the (displacement-controlled) testing processes, which normally occurs when dislocation loops move out of the pillar body, creating high local strain. In load-controlled tests this manifests as strain bursts on the stress-strain response, as those observed previously in small scale single crystal tests [57,58]. In the present case, this phenomenon is deemed a result of highly planar and rapid motion of dislocations on the prismatic plane, which is supported by the literature where prismatic slip in zirconium was found to be controlled by the interactions between mobile dislocations and obstacles such as oxygen impurities, while the Peierls friction forces are negligible [52]. Hence, it is not surprising that stress drops are significantly suppressed when the hydride packets are present, since the hydrides may retard the motion of dislocations gliding on the matrix prismatic plane as suggested by the deformation behaviour of the hydride-containing pillars.

In the case of fine grain material where the hydrides mainly decorate the grain boundaries, local shear along the hydride-matrix interfaces remains a significant path for accommodating plastic deformation around the hydrides that locate close to the plane of maximum shear for the stress state. Meanwhile, the fact that slip bands in the matrix are routinely arrested at the hydride-matrix interfaces indicates that the δ-hydride is intrinsically stronger than the α-Zr matrix, in agreement with observations in Ref. [59]. These two observations in the macroscale tensile tests are both consistent with the results of the micropillar compression tests, suggesting that the underlying mechanisms introduced above can operate regardless of specimen size, number of grains within a specimen, loading mode (tensile vs. compressive) and hydride type (in terms of nucleation sites).

It was reported previously [60] that regions with high local shear strain are potential initiation sites for ductile fracture of metallic materials. In the present case, the maximum shear strain values at the sheared hydride-matrix interfaces (as marked with the green arrows in Figure 9) are found to be in the high value region for each set of data, indicating that the presence of the hydrides could accelerate the degradation of the bulk material through promoting strain localisation at the phase interfaces and this may have contributed to the relatively weaker mechanical reponse of the hydride-containing sample (Figure 8). To quantify this, the frequency distribution of the maximum shear strain for the data for map ③ in Figure 9 is plotted in red in Figure 11(a). For comparison, the frequecy distribuion for the data obtained in a similar-sized region on the as-received fine grain Zircaloy-4 sample deformed to the same displacement to that for stage ③ in Figure 8 is plotted in blue. DIC analyses for the as-received sample were carried out using the same parameters as those employed for the hydride-containig sample introduced in section 2.2, and the maximum shear strain field plots for the as-received specimen are given in Figure S1, Supplementary material.

In Figure 11(a), the maximum shear strain (x-axis) is presented in the form of multiples of geometric mean value for each of the two datasets respectively, in order for direct comparison with eliminated inflence of the difference in the overall strain. The distributions shown in Figure 11(a) indicate that they are skew or log-normal. A probability plot, fitted to a log-normal distribution is shown in Figure 11(b). These plots indicate that both strain plots can (broadly) be described as log-normal but with either the superposition of two log-normal distribtuions, or the data presents with a log skew. Examination of the histogram presented in Figure 11(c) indicates that there is a slight left skew (i.e. a greater than expected number of observations with lower values of the normalised strain). For the peak of the two distributions, towards $\log_{10}$(normalised maximum shear strain) = 0, the frequency of observations for the as-received Zircaloy-4 sample is higher than than the hydride-containing sample. This indicates that the hydride-containing sample exhibits more heterogenous in plane shear strain, i.e. slip. This is confirmed through examination of the right of the distribution (i.e. those points with $\log_{10}$(normalised maximum shear strain) > 0, which indicates that there are substantively more areas with higher strains than the mean in the hydride-containing dataset, as well as the fact that the gradient of the log-normal probability plot for the hydride-containing sample is more shallow (Figure 11(b)). This aligns with our observation that strain is localised at the sheared hydride-matrix interfaces.

These observations can be clarified through examination of the strain field plots. Magnified maximum shear strain maps of the region within the dashed box on map ①, Figure 9 are shown in Figure 12, along with a crystal orientation map of this region showing the crystal orientations as well as the positions of the grain and phase boundaries. Noticeably, a few bands are present within the hydride packet on map ① (as marked with the yellow arrows) and the local maximum shear strain at these bands increased in magnitude progressively upon further plastic deformation (maps ② and ③). The stepwise increase of the local shear strain associated with these bands indicates that they are slip bands rather than twins, as otherwise a characteristic shear strain value that does not change over total displacement would be observed. Meanwhile, the number of slip bands in the hydride increased with the increase of the applied displacement. On map ③, slip bands can be observed throughout the hydride packet in grain *A* (the slip bands in the hydride become more clearly visible when the strain field is plotted in the form of $\log_{10}$(normalised maximum shear strain), and an example is shown in Figure S2, Supplementary material). These slip bands are not all parallel to each other but have two different sets of orientations, indicating the activation of multiple slip systems within this specific hydride packet.

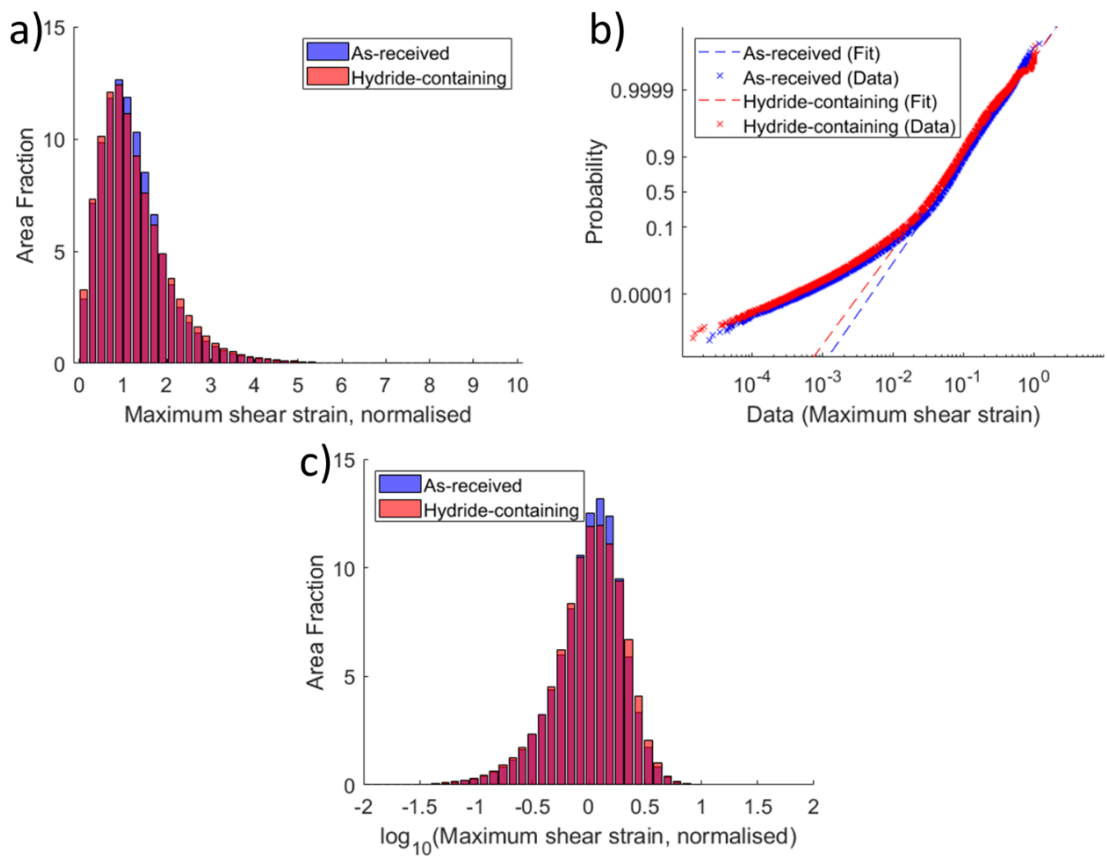

Figure 11  Analysis of the frequency distributions of the maximum shear strain (normalised against the geometric mean maximum shear strain value for each dataset) from HR-DIC data for map ③ in Figure 9 (red) and for an as-received specimen deformed to a similar displacement to that for map ③ (blue). (a) plotted as a histogram, revealing a ~log-normal distribution; (b) testing the log-normal distribution using a probability plot; (c) histogram plotted on a $\log_{10}$ scale.

Moreover, although the hydride in which slip occurred formed in grain *A*, quite a few slip bands within it are observed to interconnect with some of the slip bands in the matrix in grain *D* (as highlighted by the yellow arrows on map ③, Figure 12) and these two sets of interconnected bands have similar orientations on the strain maps. It is therefore inferred that slip within grain *D* triggered the slip observed in the hydride, and this implies that for an intergranular hydride, the initiation of interior slip can occur as a result of slip in one of its neighbouring grains but not necessarily its parent grain.

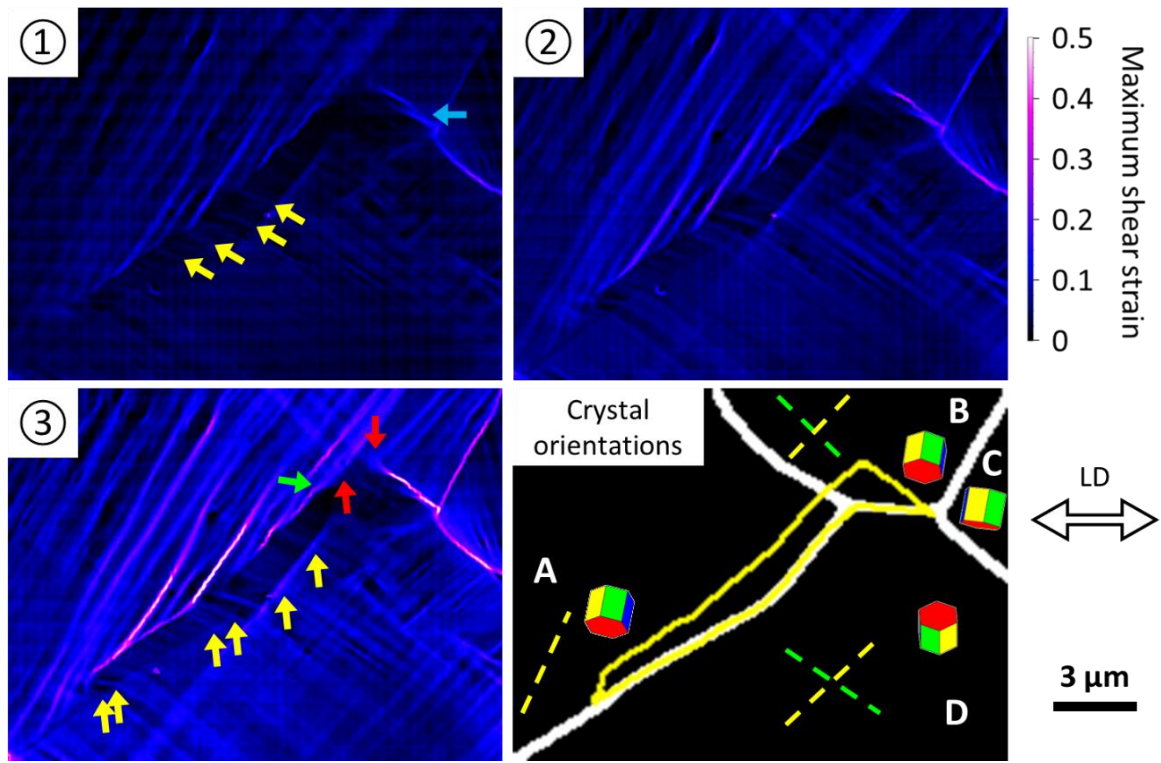

Figure 12 Magnified maximum shear strain maps of the region highlighted with the dashed box in Figure 9①, and crystal orientation map showing the crystal orientations as well as the grain and phase boundaries (grain boundaries are plotted in white and phase boundaries in yellow). LD denotes loading direction. Within each grain on the crystal orientation map, the dashed lines denote the surface trace orientations of the crystal planes drawn in the same colours on the unit cells correspondingly, and the hydride is outlined in yellow.

In grain *D,* two sets of slip bands that are nearly perpendicular to each other can be seen on the strain maps. Slip trace analysis shows that they are due to the activation of two $<a>$ prismatic slip systems (see the crystal orientation map – the dashed lines denote the surface trace orientations of the crystal planes drawn in the same colours on the HCP unit cell correspondingly). For these two sets of slip bands, one of them (the one parallel to the green dashed line in grain *D* on the crystal orientation map) is nearly perpendicular to the hydride-matrix interface on the grain *D* side, while the other (the one parallel to the yellow dashed line on the crystal orientation map) is closely-oriented to the hydride-matrix interface. The former likely resulted in the plastic slip within the hydride as introduced above, while the latter does not evidently have such effect. A broad and intense slip band in grain *D* and parallel to the yellow dashed line is observable on map ③ that extends from the hydride-matrix interface at the point where the rightmost yellow arrow is pointing, to its bottom left direction. The history of this slip band (maps ① and ②) implies that it initiated along the hydride-matrix interface and then propagated into the matrix in the form of $<a>$ prismatic slip. Similar observation is made in grain *B* where the shear event at the phase interface marked with a blue arrow on map ① triggered $<a>$ prismatic slip in the matrix towards its top left direction (maps ② and ③). It is worth noting that this interface shear-induced matrix plastic slip was also observed in one of the micropillars (Figure 4(c)), though the slip in the matrix being $<a>$ basal type due to the orientations of the pillar/hydride.

In grain *A*, slip bands of $<a>$ prismatic type can be observed on map ①, Figure 12. Some of the slip bands appear to be wavy presumably because of dislocation cross-slip. As the plastic deformation progressed, localised shear along the hydride-matrix interface on the grain *A* side was initiated at those points where the slip bands in the matrix intersect the phase boundary (map ②) and the local shear strain at the interface increased in magnitude with further deformation (map ③). Similarly, in grain *B*, $<a>$ prismatic slip in the matrix (slip bands parallel to the yellow dashed line in grain *B* on the crystal orientation map) resulted in localised shear at the hydride-matrix interface, as marked with the green arrow on map ③.

In the present case, the sample was pulled along its RD and therefore the majority of crystallographic slip in the matrix is of $<a>$ prismatic type. Meanwhile, in contrast to the intragranular hydrides in the pillars which decorate the matrix $\{10\bar{1}7\}$ planes, the habit planes of the intergranular hydrides in the macroscale tensile specimen can vary depending upon the local microstructural-crystallographic relationships [6] and this may have led to the difference in slip-hydride interaction modes between these two types of hydrides. For example, for the intragranular hydrides, the interface shear is always related to $<a>$ basal slip and the blocking of matrix shear to $<a>$ prismatic slip since the hydrides are close to the $\{0001\}$ basal plane and nearly perpendicular to the $\{10\bar{1}0\}$ prismatic planes. However, for the intergranular hydrides, $<a>$ prismatic slip alone can result in both types of slip-hydride interactions as observed in the DIC results. In general, it was noticed that the geometrical relationship between the slip band in the matrix and its neighbouring hydride determines the active interaction mode between the two, with slip bands in the matrix that are closely-oriented to the adjacent hydride-matrix interfaces triggering localised interface shearing and vice versa, and those nearly perpendicular to the adjacent hydride-matrix interfaces either leading to plastic slip within the hydrides or getting arrested at the interface.

The observation of slip bands within the hydrides in the fine grain material also proves that the formation of slip bands within the hydrides in the micropillars are not due to the size effect in micromechanical testing, where it was found previously that when the pillar size reduces to critical values, plasticity is favoured over fracture in brittle ionic crystal materials such as MgO [61,62].

Apart from hydride embrittlement, neutron irradiation in nuclear reactor cores also has an embrittling effect on Zr alloy fuel claddings [63] through the highly anisotropic formation of dislocation loops [64]. The extensive loop formation may arrest dislocation glide. This along with the elevated point defect populations after irradiation which may facilitate dislocation climb may influence the behaviour of dislocations near hydrides. In the future, the characterisation techniques employed in the present work can be applied on irradiated Zr alloys and the results will likely have even more direct industrial implications.

It is worth noting that although crystal slip is observed within the hydride packet shown in Figure 12, most of the hydrides in Figure 9 are free from noticeable interior deformation. For instance, some of the slip bands in the matrix in grain *B*, Figure 12 were arrested at the hydride-matrix interfaces (marked with the red arrows on map ③) without causing plastic slip in the hydride within this grain. This means that significant local stress can potentially build up at those points at the hydride-matrix interface where the slip bands are terminated. Even if local stress is relieved through slip within the hydrides, since the matrix crystal orientation on the other side of the hydride is normally different to where the incoming slip bands initiated in fine grain polycrystalline materials, the slip bands in the hydride will likely stop when they reach the hydride-matrix interface ahead of them and potentially lead to local stress concentration. This local deformation incompatibility as well as the shear at the hydride-matrix interface as introduced above are likely two significant early-stage microscale mechanisms that may collectively give rise to local microvoid formation and/or cracking, which are widely-believed precursors to the ultimate failure of Zr alloy fuel cladding materials. The experimental results in this work suggest that the origin of hydride embrittlement in Zr alloys may not merely be the brittle nature of the hydride phase alone, while the occurrence of plastic slip within the hydrides and the contributions from the strain/stress localisation at the hydride-matrix interfaces built up early in the plasticity regime should also be taken into consideration when understanding the failure processes.

## 5. Conclusions

The microscale mechanical interactions between δ-hydrides and plastic slip in Zircaloy-4 under stress and their effects on the mechanical performance has been studied using *in situ* SEM micropillar compression tests of single crystal specimens and *ex situ* HR-DIC macroscale tensile tests of polycrystalline specimens. The following conclusions can be drawn:

1. The room temperature CRSS values for $<a>$ basal and $<a>$ prismatic slip systems in Zircaloy-4 were measured to be around 150 MPa and 110 MPa, respectively.

2. Compared to the discrete and localised $<a>$ prismatic slip bands, the $<a>$ basal slip bands are broader and more homogeneous, presumably due to the occurrence of cross-slip. Easy and highly planar $<a>$ prismatic slip leads to significant stress drops on the stress-strain responses of the micropillars.

3. Shear along the hydride-matrix interface is favoured over $<a>$ basal slip for pillars with intragranular hydride packets sitting ~45° to the loading axis, leading to strain localisation. The interface shear can in turn lead to local $<a>$ basal slip between hydride packets.

4. Intragranular hydride packets sitting ~parallel to the loading axis can retard the propagation of $<a>$ prismatic slip bands, and this exhibits on the stress-strain responses as increased yield stress and hardening rate, as well as decreased number of stress drops. This effect of hydrides on matrix slip is insensitive to hydride thickness.

5. $<a>$ prismatic slip can trigger slip within the hydride and this may be achieved through the initiation of $<110>$-type shear in the hydride by $<a>$-type shear in the matrix due to the hydride-matrix OR.

6. Matrix slip on the planes close in orientation to the adjacent hydride-matrix interfaces can lead to intense local shear along these interfaces and vice versa, while slip on the planes nearly perpendicular to the adjacent interfaces can either get arrested at the interfaces or initiate plastic slip within the hydride. Hence, slip within intergranular hydrides can be initiated by slip in their neighbouring matrix grains which may not necessarily be their parent grains.

7. The presence of δ-hydrides results in enhanced strain localisation in Zircaloy-4 upon plastic deformation due to localised shear at the hydride-matrix interfaces. This along with the potential stress concentration at where the slip bands in the matrix are arrested at the phase boundary may act as microscale precursors to hydride-induced failure of Zr alloy fuel cladding materials.

## Acknowledgements

The authors thank Prof. Fionn Dunne for useful discussions. TBB and SW acknowledge support from HexMat (EPSRC EP/K034332/1) and MIDAS (EPSRC EP/SO1720X), as well as funding from Rolls-Royce plc through Rolls-Royce Nuclear UTC at Imperial College London. FG and SW acknowledge funding from MAPP (EPSRC EP/P006566/1). TBB thanks the Royal Academy of Engineering for funding his Research Fellowship. The FEI Quanta SEM used was supported by the Shell AIMS UTC and is housed in the Harvey Flower EM suite at Imperial College London.

**Supplementary material**

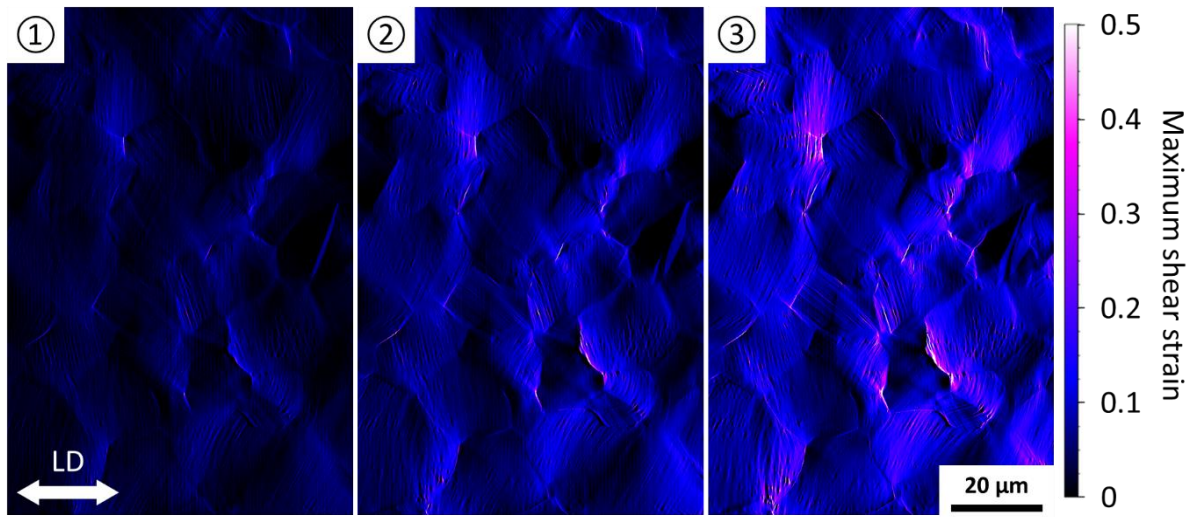

Figure S1 Maximum shear strain maps of the DIC-ROI on the as-received tensile specimen during plastic deformation. Maps ①, ② and ③ were obtained after unloading from the same displacement values as the three points marked with the crosses on the stress-displacement curve in Figure 8 correspondingly. LD denotes loading direction.

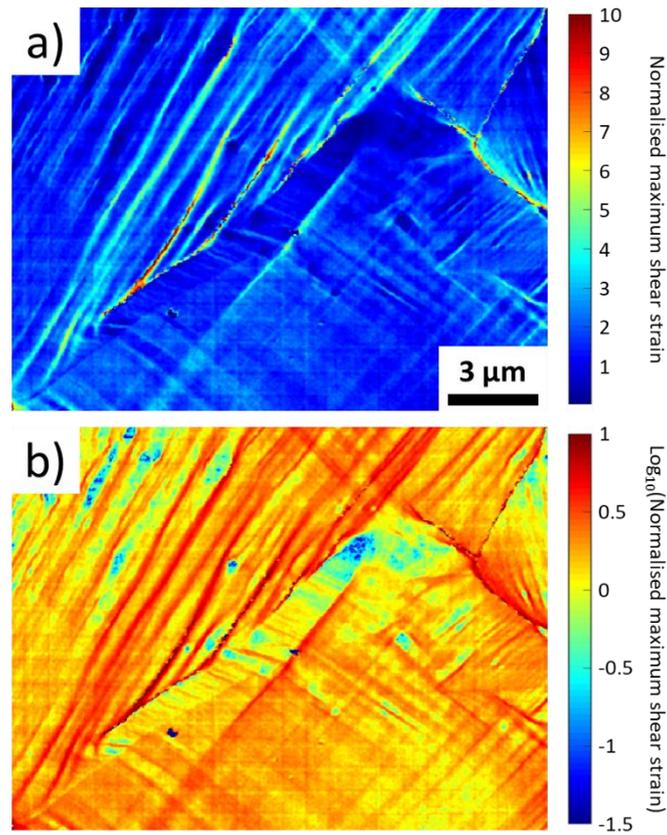

Figure S2 Maximum shear strain maps of the same region as that for Figure 12 and obtained at stage ③ in Figure 8, plotted in the forms of (a) normalised maximum shear strain (against the geometric mean value for the dataset for the hydride-containing specimen) and (b) $\log_{10}$(normalised maximum shear strain).